\documentclass[final]{siamltex}

\usepackage{color}
\usepackage{epsfig,psfrag,amsmath,amssymb,latexsym}
\usepackage{amscd}
\usepackage{amsfonts}
\usepackage{enumerate}
\usepackage{graphicx}
\usepackage{hyperref}
\usepackage{bbm}
\usepackage{mathrsfs}
\usepackage{stmaryrd} 
\usepackage{yhmath} 
\usepackage{tikz}
\usepackage[utf8]{inputenc}
\usepackage{boondox-cal}
\usetikzlibrary{arrows,decorations.markings}
\usetikzlibrary{positioning}
\usetikzlibrary{shapes.multipart,positioning,fit}
\usetikzlibrary{calc,through}
\usepackage{mdframed}

\usepackage{graphicx}
\graphicspath{{./A_Figuras_Latex/}}

\newcommand{\RR}{\mathbb{R}}

\newcommand{\PP}{\mathbb{P}}

\numberwithin{equation}{section}

\title{\textbf{IRT Scoring and the principle of consistent order}}

\begin{document}

\author{Nancy Lacourly\thanks{Center for Mathematical Modeling CMM,  UMI-CNRS 2807.
Facultad de Ciencias F\'isicas y Matem\'aticas and DEMRE,
Universidad de Chile, Beauchef 851, Santiago, Chile e-mail: nancy.lacourly@ariel.cl}
\and Jaime San Mart\'in\thanks{Departamento de Ingenier\'ia Matem\'atica \&
Center for Mathematical Modeling CMM,  UMI-CNRS 2807.
Facultad de Ciencias F\'isicas y Matem\'aticas, 
Universidad de Chile, Beauchef 851, Santiago, Chile. e-mail: jsanmart@dim.uchile.cl}
\and M\'onica Silva\thanks{.... PUC. e-mail: msilvara@uc.cl}
\and Paula Uribe\thanks{Center for Mathematical Modeling CMM,  UMI-CNRS 2807.
Facultad de Ciencias F\'isicas y Matem\'aticas, 
Universidad de Chile, Beauchef 851, Santiago, Chile. e-mail: puribe@dim.uchile.cl}}

\maketitle

\begin{abstract}
IRT models are  being increasingly used worldwide for test construction and scoring. The study examines the practical implications of estimating individual scores in a paper-and-pencil  high-stakes test using  2PL and 3PL models, specifically whether the principle of consistent order holds when scoring with IRT. The principle states that student A, who answers the same (or a larger) number of items of greater difficulty than student B, should outscore B. Results of analyses conducted using actual scores from the Chilean national admission test in mathematics indicate the principle does not hold when scoring with 2PL or 3PL models. Students who answer more items and of greater difficulty may be assigned lower scores. The findings can be explained by examining  the mathematical models, since estimated ability scores are an increasing function of the accumulated estimated discriminations for the correct items, not their difficulty.  For high stakes tests the decision to use complex model should therefore be a matter of serious deliberation for policy makers and test experts, since fairness and transparency may be compromised.  

\end{abstract}

\section{Introduction}

The ultimate purpose of testing is to assign a ``score'' to an examinee that reflects the examinees 
level of attainment of a skill measured by the test (Hambleton, Swaminathan \& Rogers, 1991, p.77). 
The scoring procedure enables the scorer to evaluate the performance in a specified domain (Standards, 2014). 

In any test, but particularly for high-stakes tests, scoring specifications need to be spelled out clearly. These 
should include whether test scores are simple sums of item scores, involve differential item weighting of items or 
sections and whether they are  based on IRT.  If an IRT model is used, specifications should indicate 
the form of the model, how model parameters are to be estimated, and how model fit is to be evaluated (Standards, 2014).
 
In recent decades, IRT models have gained momentum and are being widely used in modeling item responses and scoring' in 
 educational tests.  As in the U.S., many testing  programs around the world have adopted IRT models or are 
 evaluating a transition from classical measurement methods to IRT. Probably the greatest advantage of CTT 
 scoring is that the simple summed score is the most transparent for the general public and test takers. 
 However, the two statistics that form the cornerstones of many classical CTT analyses 
 (item difficulty and item discrimination) are group dependent. Additionally, scores obtained by CTT are test 
 dependent and comparability of results obtained from different forms of the same test is inappropriate (Hambleton \& Jones,  1993).
 
Among IRT methods, historically, the Rasch model has been more widely applied, primarily due to its robustness 
 to sample size and estimation requirements, relative ease in implementation, and straightforward score 
 interpretations for test users. However, in recent years a growing number of assessment programs have opted 
 for more complex IRT models to address additional factors besides item difficulty, such as item discrimination 
 and guessing (Tong \& Kolen, 2010). 
Unlike the more complex models, in the Rasch model scoring is straightforward  because in the statistical definition 
 of the model, the total score based on item responses is a sufficient statistic of the persons underlying ability and 
 results in a single estimate. With the more complex models, this is not the case. Tong and Kolen (2007) showed 
 that the choice of different estimators produced score distributions with different characteristics, and the differences 
 were not minor in some cases.

It is important to distinguish between the use of IRT models for test construction and scoring.
Indeed, decisions as to which IRT model should be used 
 (i.e. Rasch, 2PL or 3PL) and the choice of estimators in a testing program can affect test score distributions 
 (Tong and Kolen, 2010). In other words, different approaches (e.g. a one-parameter model versus a two-or 
 three-parameter model, or Bayesian versus ML estimates) can result in different  item and score estimates. 
 
In practice, this means that the choice of an IRT model and its underlying mathematics 
will influence student scores (Anderson, 1999). 
Experts warn of potential problems in using complex IRT models to estimate ability. 
In addition, Hambleton , Swaminathan and Rogers (1991) report that  for some peculiar or aberrant response 
patterns, the likelihood functions may fail to be properly estimated when using the 3 parameter model.  
Such aberrant response patterns can occur in situations where examinees answer some relatively difficult and 
discriminating items correctly and fail some of the easier items. 

The choice of an IRT model involves, in part, philosophical considerations such as whether the data 
should fit the model or vice versa as well as the application context such as sample size, instrument 
characteristics, assumption tenability and political realities, among others (de Ayala, 2009). The most widely used  
model -- Rasch or 1PL -- is the most restrictive, but a number of studies investigating its use when it misfits 
show that it yields reasonably invariant item parameters and ability estimates (Forsyth et at. cited by de Ayala, 2009). 
Complex IRT models (2PL and 3PL) have detractors (Cliff, 1996; Michell, 1999, among others). In particular, 
Wright (1999) recommends the use of Rasch models and suggests that additional parameters are 
unnecessary, wipes out additivity and that crossing item characteristic curves (ICC) blurs the construct 
that is being measured. Specifically, Wright cautions against the use of 2PL and 3PL models that cause the hierarchy 
of relative item difficulty to change at every ability level.

The choice of model becomes particularly relevant in high-stakes testing programs, since the assignment of scores 
using complex IRT models can have practical implications on the resulting scores. Those in charge of testing agencies 
have a major responsibility in deciding which model to use and in justifying their choice. Summed scores are more 
transparent and have a straightforward interpretation for test users.  The public needs to be educated as to the 
advantages and disadvantages of IRT models, particularly if 2PL or 3PL models are used.

\noindent{\bf Purpose}

The purpose of this paper is to examine the practical implications of estimating individual scores in a high-stakes test using 
1PL, 2PL or 3PL models. The study focuses on the effects for individual test takers of applying different models using 
actual test data from 2016 Chilean university admission process that relies heavily on test scores. The only selection criteria employed by the centralized admission system are test scores, high-school grades, and student class-rank in high-school.  
We explore whether differences in item calibration and scoring based on different models have an impact on the relative 
standing of applicants and their actual test scores, which can ultimately affect the admission decision. 

The findings should be useful to inform policy decisions as to which model should be used when reporting test scores. 
Fairness and transparency issues should also weigh on the decision to use one or another, particularly when 
scores bear high-stakes for individuals. A key issue is whether the estimated ability between students is consistent 
with the number and difficulty of correct items responded by each individual. 

In this article we explore the {\bf principle of consistent order} (PCO) a person that answers the same number of correct items 
with greater difficulty than another
should attain a higher score. Specifically, let us assume that two students {\it A} and {\it B} take the same 
pencil-and-paper test and answer correctly the  same number of questions. When we rank order the difficulty of the 
questions responded by both students in a descending order we notice that all the questions answered by  
{\it B}  are more difficult than those answered by {\it A}. 
Thus, we would expect that  {\it B} should obtain a higher score than {\it A}.  
Would that hold true using 2PL or 3PL models?. 
The paper examines whether the PCO holds when using 2PL and 3PL models. In the 1PL and classical theory, student $A$
and $B$ will have at least the same rank order, and the PCO is not violated.

To understand the PCO let us consider the following scenario. 
For each student $j$  consider the difficulties  $(b_i: i\in C_j)$ of the items correctly responded by him/her. 
We order these difficulties in decreasing fashion:
$$
\mathcal{V}_j=\left(b^j_{(1)}\ge  b^j_{(2)}\ge \cdots \ge b^j_{(n_j)}\right)\,,
$$
where $n_j$ is the number of correct items for student $j$ and we add an extra index $j$ because this 
order depends on the student $j$: $b^j_{(1)}$ is the difficulty of the most difficult item answered correctly by student $j$, 
$b^j_{(2)}$ is the difficulty of the second most difficult item answered correctly by student $j$, and so on.

\bigskip

We say student $k$ is {\it weaker} than student $j$, which we denote by $k \prec j$, if the following conditions hold:
first $n_k\le n_j$ and for $\ell=1,\cdots,n_k$
$$
b^k_{(\ell)}\le b^j_{(\ell)}\,,
$$
with at least one strict inequality.

\medskip

That is, $j$ answered correctly more items than $k$ and the most difficult item answered by $j$ is more difficult than the
most difficult item answered by $k$; the second most  difficult item answered correctly by $j$ is also more difficult than the second 
item correctly answered by  $k$ and this holds for all the $n_k$ items answered correctly by student $k$. Thus, any 
reasonable score should put student $j$ ahead of student $k$.

\medskip

The main question here is: Is it possible that there are two students $k$ {\it weaker} than $j$ ($k\prec j$) 
such that $\theta_j<\theta_k$? Namely, even though $k$ is a {\it weaker} student than $j$, $k$ has a 
higher estimated ability score.
In what follows, we will say that $k$ {\bf dominates} $j$, or that $j$ is {\bf dominated} (i.e. disadvantaged) by $k$
if $k$ is {\it weaker} than $j$ but nevertheless has a higher estimated ability.

\medskip

If $\theta_j<\theta_k$ then the estimated ability is not an increasing function of the difficulties of the correct items as 
one would expect. We will check whether this holds in 2PL and 3PL models using actual data from the Chilean national selection test (PSU 2016).

\section{Method}

\subsection{Data}  We use the data base of student responses to
the national university admission test in Mathematics in 2016 (PSU). The PSU is 
a paper-and-pencil  multiple-choice high-stakes test  with $75$ items 
dichotomously scored.  The test has 4 forms and was taken by 252,745 students. 
Form 1 was selected to conduct the study, which was taken by $N=63,498$ students. 

Scoring in PSU has traditionally been done using CTT. However, a recent evaluation of Chilean university 
admission tests conducted by independent experts  recommended the adoption of IRT methods for test 
construction and to equate test forms across administrations (Pearson, 2013). Although the authors of the 
Pearson Report did not specifically recommend a departure from the traditional sum of item scores to IRT 
scoring, agency experts are debating whether to switch to 2PL or 3PL IRT scoring.

\subsection{Procedure} 
In order to examine the PCO we associate to each student $k$ the following variables 
\begin{enumerate}[(1)]
\item $n_k=$number of correct items for student $k$;

\item $\widehat{\theta}_k=$ estimated ability of student $k$
\item $Di\!f_k=$ vector of (estimated) difficulties of items answered by student $k$ in descending order, where the incorrect
items are filled with a large negative number (-100 for example). We have $Di\!f_k\in \RR^N$, where
$N$ is the number of valid items in the test.
\end{enumerate}

\medskip
For every student $k$, we search for all possible students $j$ that are dominated relative to $k$, as follows,
$$
S_k=\{j:\, n_k\le n_j,\, Di\!f_k \preceq Di\!f_j,\, \widehat{\theta}_j < \widehat{\theta}_k\}
$$
that is, $j$ has answered more correct items than $k$ and  $j$ difficulties are larger than those of  $k$, but $j$ has smaller
estimated ability than $k$.

\subsection{Calibration and scoring} 
The IRT analyses were run in R using MML estimation of the parameters. We
estimate student's ability using both EAP and WLE.  
To check if the principle of strong order holds, we used a MATLAB 
program (see Appendix XXXX for its flow diagram).

\section{Results}
\label{sec:Numerical} 
In this section we describe the main findings for 2PL and 3PL. We found its use resulted in an inconsistent ranking among students
counter to the PCO.

\subsection{Dimensionality Analysis} 
\begin{figure}[h]
\centering
\includegraphics [width=10cm,height=6cm]{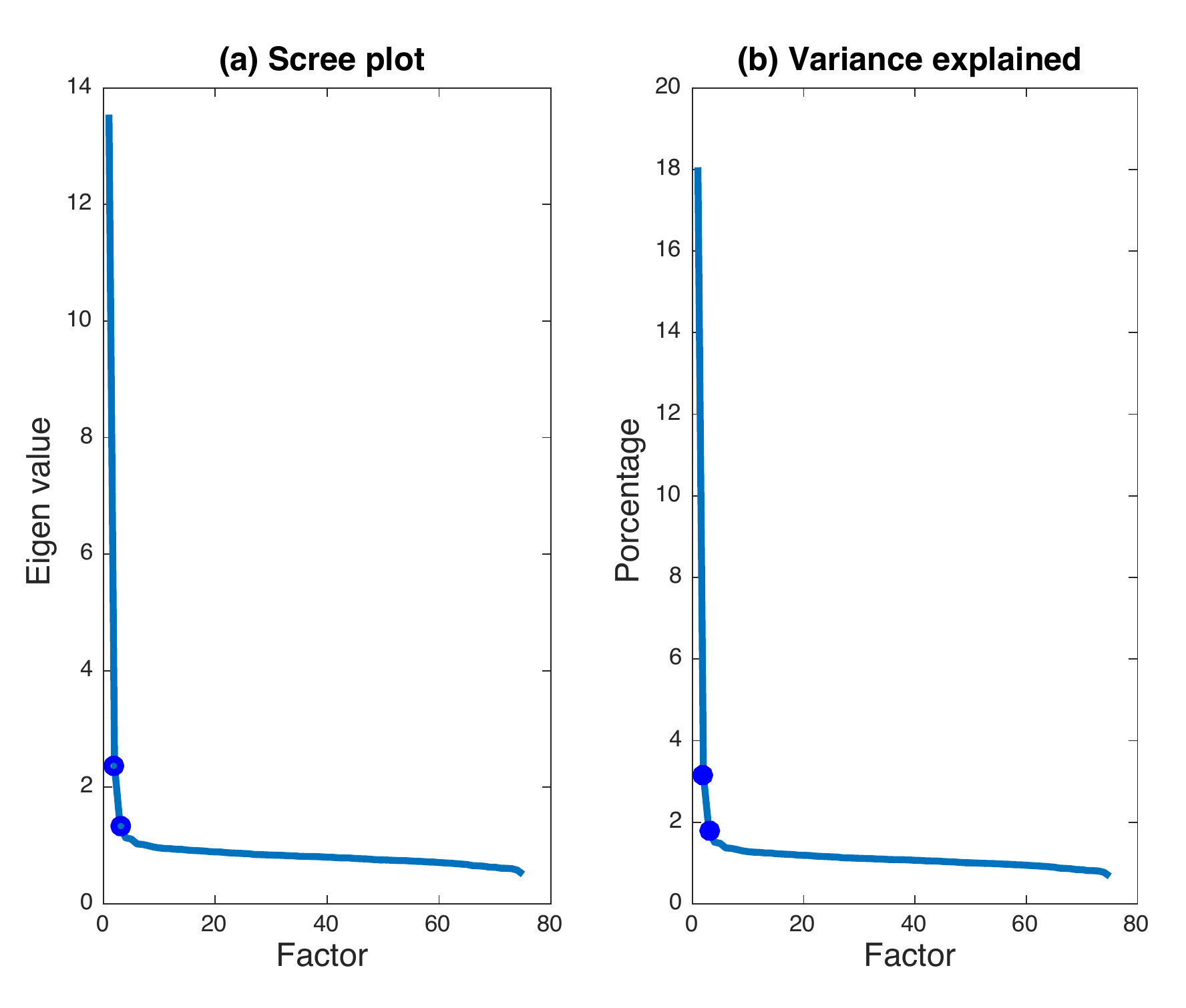}
\vspace{-0.5cm}
\caption{Relation between discrimination and difficulty and ability scores}
\label{fig:1}
\end{figure}

The data were factor analyzed to assess the tenability of the unidimensionality assumption prior to estimating scores using IRT models. 
This assumption appears tenable due to the presence of a dominant factor that explains 18\% of the variance. The second and third factors, albeit statistically significant, 
explained very little of the remaining variance  (figure \ref{fig:1}(b)). The scree plot confirms the presence of one strong factor  (figure \ref{fig:1}(a)).

\subsection{Examination of PCO}

\begin{figure}[h]
\centering
\includegraphics [width=12cm,height=6cm]{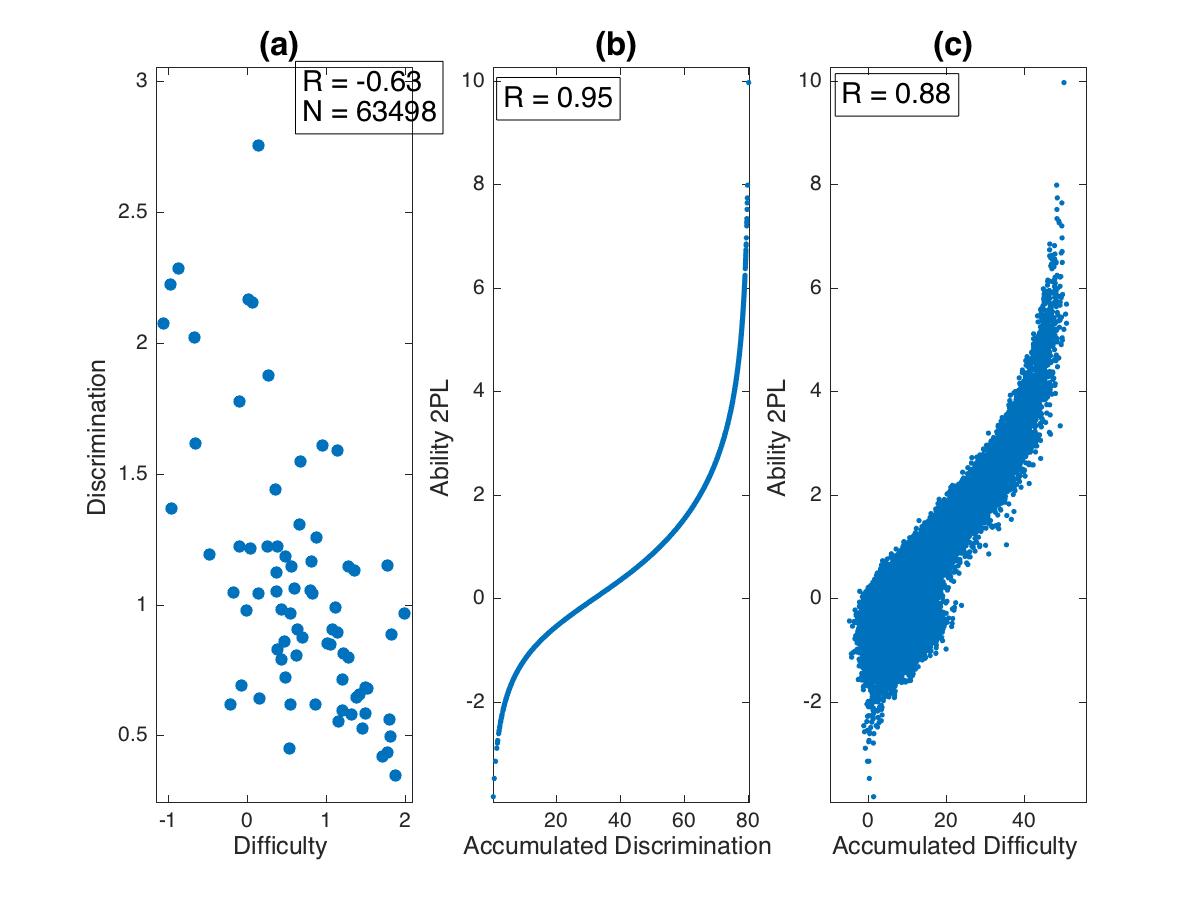}
\vspace{-0.5cm}
\caption{2PL model: (a) Difficulty v/s discrimination, (b) Ability v/s Accumulated discrimination  and (c) v/s Accumulated difficulty}
\label{fig:2}
\end{figure}

The relation between difficulty and discrimination in the 2PL model, Figure \ref{fig:2}(a), shows the absence of a monotonic positive association between
difficulty and discrimination. Even for small (positive) values of the difficulty, these two variables are not increasingly related. 
Figure \ref{fig:2}(b) shows that ability is an increasing function of the accumulated discrimination of correct items, 
as the theory predicts and not an increasing function of the difficulties of correct items (figure  \ref{fig:2}(c)). 

\begin{figure}[h]
\centering
\includegraphics [width=10cm,height=9cm]{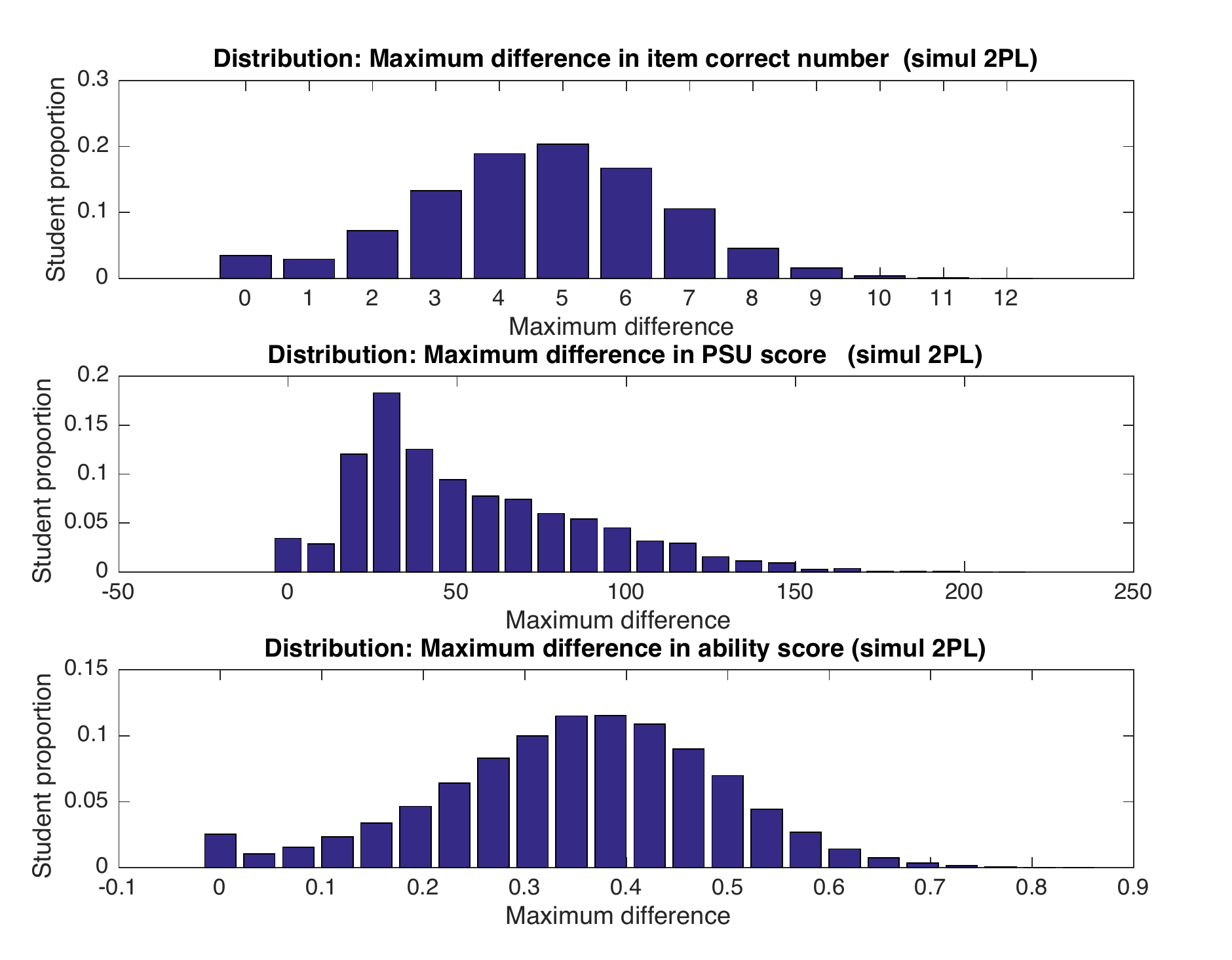}
\vspace{-0.5cm}
\caption{Maximum observed differences in 2PL: correct items, PSU scores and ability scores}
\label{fig:4}
\end{figure}

In Figure \ref{fig:4} we report  histograms for the differences in the number of correct items, PSU scores and
estimated ability between the dominating and dominated students. Histogram (a) shows the maximum 
number of additional correct items responded by dominated students, which ranges from 0 to 12.
Histogram (b) shows the same difference in actual PSU scores. Considering that PSU scores range from 150 to 850 points with
an average of 500 and a standard deviation of 110, the observed differences are substantive with over 50\% of these exceeding 
one half of a standard deviation in PSU scores. Finally, histogram (c) shows this difference expressed as 2PL ability scores, in a scale
that ranges from -2.53 to 4.23.

Table \ref{Tab:1} summarizes the findings related to PCO. The first column represents the selected categories according to the 
number of items responded correctly by students. These categories are:
10, 20, 30, 40, 50, 60, 70 and 74. The category of 75 was omitted because students with the maximum of 75 correct items 
cannot be dominated by any other student. The column labeled {\it Number of students} indicates the number of students in the given category. 
For example 2,682 students (out of 63498) correctly responded 20 items. 

Column three indicates the number of dominating students per score category, and
column four reports its respective percentage. For example, for the 20 correct items category, column four shows 
that 98 \% dominates over at least one student. 
In this category each of the 2,632 students dominate in average over 970.84 students, as reported in column five. 

Column six shows the mean number of additional correct items answered
by the dominated  students. For example, in the 20 correct items category, we have 1.43 additional correct items responded
by the dominated students. 
The maximum item difference is given in the next column. For students with 20 correct items, we observe that some dominated
student responded 10 additional correct items (i.e. 30 correct items).
The last two columns report the difference in estimated ability between the dominating and the dominated students.
In average the dominating students have 0.14 extra ability points and a maximum difference of 0.70 ability points. 

\begin{table}[h]
{\scriptsize
\begin{tabular}{|c|c|c|c|c|c|c|c|c|} \hline 
Number 	& Number	 	& Number      	&  Percen-   	&  Mean          	& Mean         	&  Maximum  	&   Mean        	& Maximum 	\\
correct 	& students	& dominating 	&  tage        	&  number       	&  items         	& items         	&  ability         	& ability  		\\ 
items	&	         	& students     	&		        &  dominated  	&  difference  	& difference  	&  difference  	&  difference 	\\
		&	         	&     			& 	             	&  students     	&                    	&                    	&                   	&           	 	\\ \hline
10 		&	911		&	876   	&	0.96		&	681.11	&	1.91		&	  9.00	&	0.11	 	&	0.67		\\
20		&	2682 	&	2632    	&	0.98		&	970.84	&	1.43		&	10.00	& 	0.14	 	&	0.70		\\
30		&	997		&	974      	&	0.98		&	300.85	&	1.69		&	10.00	&	0.11 		&	0.57		\\
40		&	584		&	554       	&	0.95		&	160.81	&      1.56 		&     	12.00	&	0.10		&	0.58		\\
50		&	385		&	366     	&	0.95		&	109.66	&      1.24 		&     	  8.00 	&	0.11		&	0.72		\\
60		&	275		&	256     	&	0.93		&        69.23	&	1.46		&	  5.00	&	0.12		&	0.61		\\
70		&	161		&	146     	& 	0.91		&	  53.62	&      0.51		&	  4.00	&	0.16		&	0.85		\\
74		&	51		&	49     	&	0.96		&	  19.29	&	0.00		&	  0.00	&	0.15		&	0.48		\\				
\hline
\end{tabular}
}
\caption{Summary of the differences for selected categories in 2PL model}
\label{Tab:1}
\end{table}
The violation of the PCO is present in all categories, from 1 to 74 correct items:   
Out of the 63,498 students that took that test, almost all 62,044 (97.7\%) dominates over someone else. 
The most extreme case is represented by a student that 
dominates over 21,054 students, which is more than a third of the total number of test takers (not in table).

In the higher end of the score spectrum (60 or more correct items) the violation of the strong order principle 
can have important consequences 
because it affects students who are likely to compete for the most prestigious slots in public universities.

\begin{figure}[h]
\includegraphics [width=14cm,height=12cm]{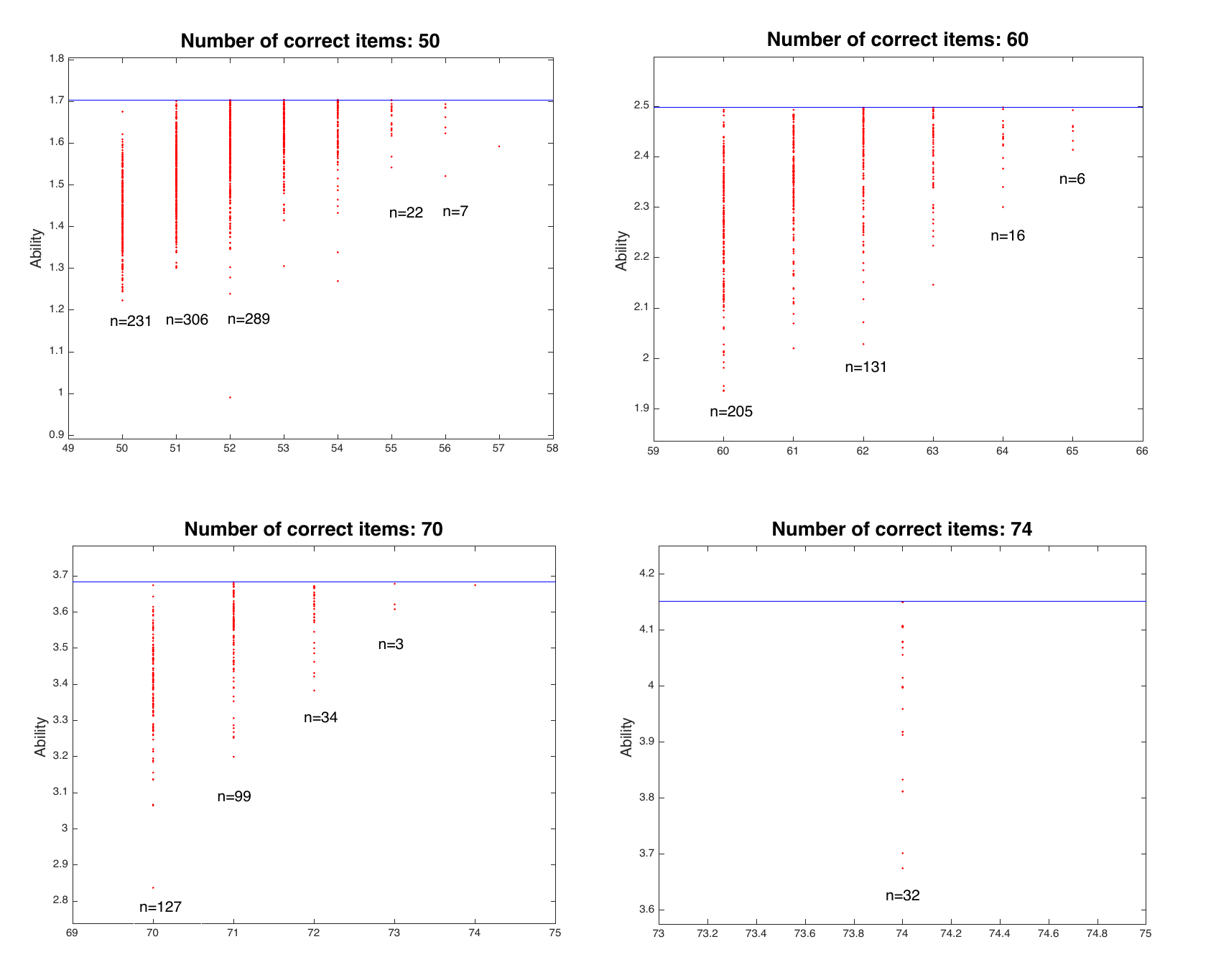}
\vspace{-1cm}
\caption{Cases 50, 60, 70 and 74 correct items, 2PL model}
\label{fig:5}
\end{figure}

In  Figure \ref{fig:5} we present selected cases in the score categories of 50, 60, 70 and 74. 
For example, the first subplot represents the set of overpassed 
students for a particular dominating student with 50 correct answers, whose ability is approximately $\theta=1.7$. 
Every line in red represents the range of estimated ability of dominated students
classified by the number of correct answers. For example the first line corresponds to the 231 dominated students
that have the same 50 correct answers. However, every dominated student answered 50 correct items 
with a higher level of difficulty than those answered by the dominator, yet all 231 students
have lower estimated ability (the lowest being 1.22).

Similar results are obtained for the  model and the corresponding figures and tables can be found in Appendix 
\ref{sec:Numerical 3PL}. 
\clearpage
\vfill\eject

\section{Discussion}
We have shown evidence using actual test data that the application of the 2PL model can result in a violation
of the principle of strong order. It is tempting to conclude that this can be due to a faulty test or to the violation of
IRT assumptions. However we replicated the analysis using an international test and found similar results. There is a mathematical explanation that points to the fact that the violation of the PCO
may be a structural problem of the 2PL model, regardless of the quality of the test and/or whether IRT assumptions
are met.
Experts have warned of potential problems in using complex IRT models to estimate ability and our study provides
empirical evidence using real test data that demonstrates that the impact is far from negligible.

We proceed to lay out the mathematical underpinnings of the 2PL model, specifically
how model parameters and model fit are estimated. An examination of the mathematical equations provide an explanation
of why the PCO is violated. We consider two estimation procedures: 
JML and MML. Although we used the MML estimation 
in the empirical analysis, we will also introduce JML because it provides a more straightforward explanation of 
the phenomenon and shows that the violation PCO regardless of the estimation method.

\subsection{The 2PL model with JML}
Assume we estimate $(\theta_j:\, j\in \mathcal{A}), ((a_i,b_i):\, i\in \mathcal{I}$) using Maximum Likelihood
estimation in the 2PL IRT model. The Likelihood function for each student $j$ is
$$
L_j=\prod\limits_{i=1}^n P_{ij}^{u_{ij}} (1-P_{ij})^{1-u_{ij}}=\prod\limits_{i\in C_j} P_{ij} \, \prod\limits_{i\notin C_j} (1-P_{ij})
$$
where 
$$
\begin{array}{l}
P_{ij}=\frac{e^{a_i(\theta_j-b_i)}}{1+e^{a_i(\theta_j-b_i)}}=P(\theta_j,a_i,b_i)\\
\\
u_{ij}=\begin{cases}1 &\hbox{ student } j \hbox{ answers correctly item } i\\
					   0 &\hbox{ otherwise}
		\end{cases}\\
\\					   
C_j=\{i\in \mathcal{I}:\, u_{ij}=1\}\, \hbox{ the set of correct items for student } j\,.
\end{array}
$$
The total likelihood $L$ is the product $L=\prod\limits_j L_j$.  Recall that $P(\theta,a,b)$ represents the 
probability that a student with ability $\theta$ answers correctly a question with discrimination $a$ 
and difficulty $b$. This function has to be increasing in $\theta$, which amounts to say that $a>0$. 
In what follows we denote by ${\bf a}=(a_i: \, i\in  \mathcal{I}), {\bf b}=(b_i: \, i\in  \mathcal{I})$.

In order to maximize $L$, we study the first order equations associated to $\mathscr{L}=\log(L)$, which are given by
{\small
\begin{eqnarray}
\label{eq:1}
&\frac{\partial \mathscr{L}}{\partial a_i}=\sum\limits_{j\in  \mathcal{A}} \, (\theta_j-b_i) 
\left[ u_{ij}  (1-P_{ij})-(1-u_{ij}) P_{ij}\right]=0\nonumber\\
&=\sum\limits_{j\in  \mathcal{A}: \, u_{ij}=1} \, (\theta_j-b_i) -\sum\limits_{j\in \mathcal{A}} (\theta_j-b_i) P_{ij}=0\\
&\nonumber\\
\label{eq:2}
&\frac{\partial \mathscr{L}}{\partial b_i}=-\sum\limits_{j\in  \mathcal{A}} \, a_i \left[ u_{ij}  (1-P_{ij})-(1-u_{ij}) P_{ij}\right]
=-a_i\left[\sum\limits_{j\in  \mathcal{A}: \, u_{ij}=1} \, 1 -\sum\limits_{j\in \mathcal{A}} P_{ij}\right]=0\\
&\nonumber\\
\label{eq:3}
&\frac{\partial \mathscr{L}}{\partial \theta_j}=\sum\limits_{i\in  \mathcal{I}} \, a_i \left[ u_{ij}  (1-P_{ij})-(1-u_{ij}) P_{ij}\right]
=\left[\sum\limits_{i\in C_j} \, a_i -\sum\limits_{i\in \mathcal{I}} \, a_i P_{ij}\right]=0
\end{eqnarray}
}Equation (\ref{eq:3}) gives the ability $\theta_j$ of student $j$ as a function of $\sum\limits_{i\in C_j} \, a_i $, which
is the accumulated discrimination of the items he/she answered correctly. We notice that the function
$$
g(\theta)=\sum\limits_{i\in \mathcal{I}} \, a_i P(\theta,a_i,b_i)=\sum\limits_{i\in \mathcal{I}} \, a_i \,\frac{e^{a_i(\theta-b_i)}}
{1+e^{a_i(\theta-b_i)}}\,,
$$
is strictly increasing in $\theta$ (because all $a_i>0$), and it is the same function for all students.  
Equation (\ref{eq:3}) is equivalent to
$$
g(\theta_j)=\sum\limits_{i\in C_j} \, a_i\,.
$$
The solution of this equation is
$$
\theta_j=g^{-1}\left(\sum\limits_{i\in C_j} \, a_i\right)\,,
$$
which is an increasing function of the accumulated discrimination (of the correct items answered by  student $j$). The important
observation is that this function is {\bf common} to all students. Then, once we compute the estimators $\widehat{\bf a}, \widehat{\bf b}$, 
using the equations above, we get  for student $j$
$$
 \sum_i \hat a_i u_{ij}=\sum\limits_{i\in \mathcal{I}} \, \hat a_i P(\theta,\hat a_i,\hat b_i)=
 \sum\limits_{i\in \mathcal{I}} \, \hat a_i \,\frac{e^{\hat a_i(\theta-\hat b_i)}}
{1+e^{\hat a_i(\theta-\hat b_i)}}=\widehat g (\theta)\,.
$$
In summary, the ability of a student is an increasing transformation of the 
accumulated (estimated) discrimination of the correct items. 

\smallskip
In the case of Rasch or 1PL models, we arrive to a similar conclusion, namely that,
the ability of student $j$ is an increasing function of 
$$
a \sum\limits_{i\in C_j} 1=a |C_j|\,
$$
where $a$ is the constant discrimination for the Rasch Model or $a=1$ in the 1PL model, and $|C_j|$
is the number of correct items for student $j$. This means, in both models the ranking of the students in Rasch
is the same as in the classical model, a result which is well established.

\subsection{The 2PL model with MML} The MML estimation considers the ability of students as a random sample
from a distribution $G(\theta)$, which is customarily assumed to belong to a family $\mathscr{G}$. For example
$\mathscr{G}$ could be the family of normal distributions with mean $0$ and (unknown) variance $\sigma^2$. Other important
case is to consider a discrete version, where one considers a priori a bounded interval, say $[-4,4]$, and {\it discretize} it in a finite
number of points $\theta_1=-4<\theta_2<\cdots< \theta_{p-1}<\theta _p=4$. In this case $\mathscr{G}$ can be identified with the
family of weights $(w_1,\cdots,w_p)\in [0,1]^p$ with the extra assumption $\sum\limits_{i=1}^p w_i=1$. In what follows we
denote by $dG(\theta)$ the measure associated to the distribution $G$.

The MML method, for the 2PL model, considers that for each student $j$ we have
$$
\PP_G(U_j=u_j \big | \mathbf{a}, \mathbf{b})=\int \frac{\exp(\theta \sum_i a_i u_{ij} -\sum_i a_i b_i u_{ij})}
{\prod\limits_i (1+\exp(a_i\theta -a_ib_i))} dG(\theta)\,.
$$  
Then the method proceeds to estimate $\mathbf{a}, \mathbf{b}$, and $G$ by maximizing the likelihood
$$
\max\left\{ \prod\limits_j \PP_G(U_j=u_j \big | \mathbf{a}, \mathbf{b}): 
\mathbf{a}\in \RR_+^n, \mathbf{b} \in \RR^n, G\in \mathscr{G} \right\}\,.
$$
Once the Maximum Likelihood estimators are obtained $\widehat{\mathbf{a}}, \widehat{\mathbf{b}}, \widehat{G}$, we consider
the distribution (posterior) of the ability for every student
$$
d\,\PP( \Theta=\theta \big | U_j,\widehat{\mathbf{a}}, \widehat{\mathbf{b}}, \widehat{G})=
\frac{1}{M(U_j  \big | \widehat{\mathbf{a}}, \widehat{\mathbf{b}}, \widehat{G})} 
\frac{\exp(\theta \sum_i \hat a_i u_{ij} -\sum_i \hat a_i \hat b_i u_{ij})}
{\prod\limits_i (1+\exp(\hat a_i\theta -\hat a_i \hat b_i))}\, d\widehat{G}(\theta)\,.
$$
Here $M(U_j  \big | \widehat{\mathbf{a}}, \widehat{\mathbf{b}}, \widehat{G})$ is a normalizing factor 
and corresponds to the marginal distribution of $U_j$ conditional on $(\widehat{\mathbf{a}}, \widehat{\mathbf{b}}, \widehat{G})$. 
Consider $T(U_j)=\sum_i \hat a_i u_{ij}$ the accumulated (estimated) discrimination of the correct items
answered by the student. Let us compute the joint distribution of $\Theta, T$
$$
\begin{array}{l}
d\,\PP( \Theta=\theta, T=t \big |\widehat{\mathbf{a}}, \widehat{\mathbf{b}}, \widehat{G})=
\sum\limits_{u: T(u)=t}  d\,\PP( \Theta=\theta, U_j=u \big | \widehat{\mathbf{a}}, \widehat{\mathbf{b}}, \widehat{G})\\
\\
=\sum\limits_{u: T(u)=t} \frac{\exp(\theta \sum_i \hat a_i u_{ij} -\sum_i \hat a_i \hat b_i u_{ij})}
{\prod\limits_i (1+\exp(\hat a_i\theta -\hat a_i \hat b_i))}\, d\widehat{G}(\theta)
=\frac{\exp(\theta t)}{\prod\limits_i (1+\exp(\hat a_i\theta -\hat a_i \hat b_i))} 
\sum\limits_{u: T(u)=t} \exp(-\sum_i \hat a_i \hat b_i u_{ij}) \, d\widehat{G}(\theta)
\,.
\end{array}
$$
Hence, we conclude 
$$
d\PP( \Theta=\theta \big | T=t, \widehat{\mathbf{a}}, \widehat{\mathbf{b}}, \widehat{G})=
\frac{1}{N(t\big | \widehat{\mathbf{a}}, \widehat{\mathbf{b}}, \widehat{G})} \frac{\exp(\theta t)}
{\prod\limits_i (1+\exp(\hat a_i\theta -\hat a_i \hat b_i))}\, d\widehat{G}(\theta)\,.
$$
$N(t \big | \widehat{\mathbf{a}}, \widehat{\mathbf{b}}, \widehat{G})$ is a normalizing factor and 
corresponds to the marginal distribution of $T$.
The distribution $d\,\PP( \Theta=\theta \big | T=t, \widehat{\mathbf{a}}, \widehat{\mathbf{b}}, \widehat{G} )$ 
has a density with respect to $d\widehat{G}(\theta)$ that can be factorized as 
\begin{equation}
\label{eq:factor}
\PP( \Theta=\theta \big | T=t, \widehat{\mathbf{a}}, \widehat{\mathbf{b}}, \widehat{G})=\exp(\theta t) h(t) c(\theta)\,.
\end{equation}
Here we think $T$ as a parameter and  $\Theta$ as the variable.
This form of the density implies that when $f$ is an increasing function then the estimator 
$$
\widehat{\theta}_j\left(t, \widehat{\mathbf{a}}, \widehat{\mathbf{b}}, \widehat{G}\right)=
\int f(\theta) \PP\left( \Theta=\theta \big | T(U_j)=t, \widehat{\mathbf{a}}, \widehat{\mathbf{b}}, \widehat{G}\right) d\widehat{G}(\theta)
$$
is an increasing function of $t$ (when $\widehat{\mathbf{a}}, \widehat{\mathbf{b}}, \widehat{G}$ are kept fixed).
In particular the posterior mean (Bayes estimator)
is an increasing function of $T$. Also the posterior median is an increasing function of $T$. 

Moreover, we shall demonstrate that the posterior mode is also increasing in $T$. For that purpose we assume 
that $\hat G$ has a density $\widehat{\mathcal{g}}$, either with respect to the 
Lebesgue measure or with respect to a counting measure.

Assume $\hat \theta(t_1)$ is the posterior mode, which
satisfies in particular for all $\theta\le \hat \theta(t_1)$
$$
\frac{1}{N(t_1)} \exp(\hat \theta(t_1)\, t_1)\, \widehat{\mathcal{g}}(\hat \theta(t_1))\ge 
\frac{1}{N(t_1)} \exp(\theta\, t_1)\, \widehat{\mathcal{g}}(\theta)\,,
$$
or equivalently
$$
\exp((\hat \theta(t_1)-\theta)\, t_1)\, \widehat{\mathcal{g}}(\hat \theta(t_1))\ge \widehat{\mathcal{g}}(\theta)\,,
$$
Consider now $t_2> t_1$. Since $(\hat \theta(t_1)-\theta)\ge 0$ we obtain
$$
\exp((\hat \theta(t_1)-\theta)\, t_2)\, \widehat{\mathcal{g}}(\hat \theta(t_1))\ge 
\exp((\hat \theta(t_1)-\theta)\, t_1)\, \widehat{\mathcal{g}}(\hat \theta(t_1))\ge 
\widehat{\mathcal{g}}(\theta)\,,
$$
and therefore for all $\theta\le \hat \theta(t_1)$ we get
$$
\frac{1}{N(t_2)} \exp(\hat \theta(t_1)\, t_2)\, \widehat{\mathcal{g}}(\hat \theta(t_1))\ge 
\frac{1}{N(t_2)} \exp(\theta\, t_2) \, \widehat{\mathcal{g}}(\theta)\,.
$$
This shows that the mode of the posterior density conditional to $T=t_2$, has to be larger or equal than $\hat \theta(t_1)$,
that is,
$$
\hat \theta(t_1)\le \hat \theta(t_2)\,,
$$
as we wanted to prove.

\bigskip

Let us recall that an interesting consequence of \eqref{eq:factor} 
is that we can produce a strong coupling: If $t_1<t_2$ then we can construct  two random variables
$\Theta_1, \Theta_2$ such that
$$
\begin{array}{l}
\Theta_r\sim \PP( \Theta=\theta \big | T=t_r, \widehat{\mathbf{a}}, \widehat{\mathbf{b}}, \widehat{G}),\, r=1,2 \\
\\
\Theta_1\le \Theta_2\,, \hbox{as random variables}\,.
\end{array}
$$
This means that from the statistical point of view the student $j$ with statistic $T(U_j)=t_1$
is fully dominated by the student $k$ with statistic $T(U_k)=t_2$.

\medskip

The interpretation of this observation is that, for the 2PL model, the ability score of the students is ranked as his/her accumulated
discrimination, that could be contrary to PCO.

\medskip

Summarizing, on the one hand  model estimated discrimination and estimated difficulty in the 2PL model are not positively
related. On the other hand, the mathematics of the model shows that the estimated ability is an increasing function of the 
accumulated estimated discriminations of the correct items, not its difficulties. 
So, a student gets a larger estimated ability if she correctly answers the 
items with larger discrimination and she can end up dominating another student that has answered correctly 
more items, which are more difficult but with smaller accumulated discriminations. Is this fair?

The answer to this question poses a dilemma for test developers: Which model should be used for scoring high-stakes tests? When using classical 
theory methods or 1 PL it is evident that the difficulty level of the items responded does not influence the score. Scores may differ between methods, 
but within each scoring method  the rank order of students is consistent: the more questions answered the higher the score, irrespective of the difficulty level of the items.

 It is tempting to adopt a more sophisticated scoring scheme amd use complex IRT models that take into account item difficulty, discrimination and guessing. However, from a 
 fairness and transparency perspective it is difficult to defend the rationale of a scoring procedure that does not respect the PCO. Before a decision is made to use 2PL or 3PL 
 for scoring purposes  it is crucial to assess its real impact in admission decisions, particularly for those at the higher end of the score spectrum and examine whether moving 
 from simpler to more complex models has a positive impact in terms of increases in the predictive validity of test scores. The use of more complex models is not necessarily 
 better and the option for a more complex model should be justified in terms of gains in  fairness, efficacy and efficiency of its use.

\vfill\eject
\section{Figures and Tables PSU 2016, 3PL}
\label{sec:Numerical 3PL}
$ $

\begin{figure}[h]
\centering
\includegraphics [width=12cm,height=7cm]{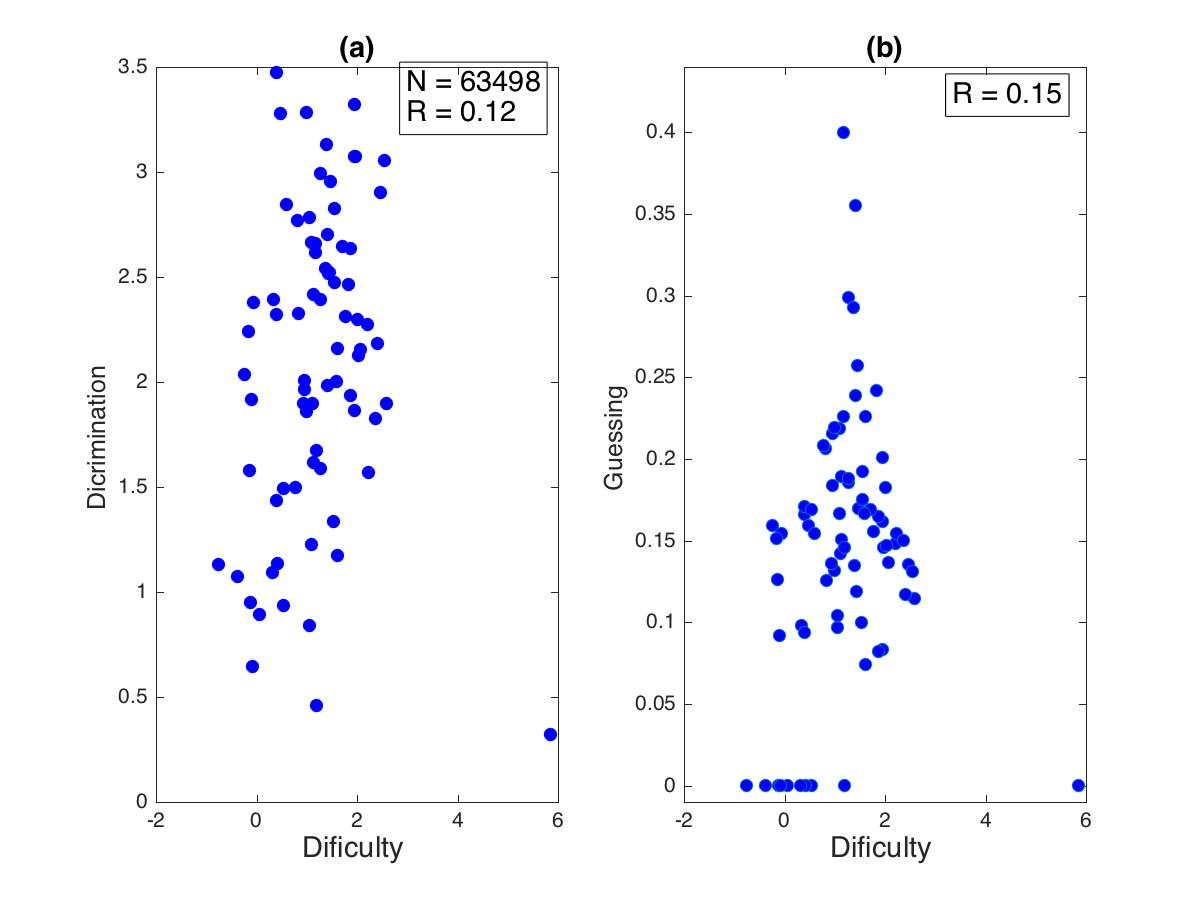}
\vspace{-0.5cm}
\caption{Difficulty v/s discrimination and guessing in 3PL}
\label{fig:7}
\end{figure}

\begin{figure}[h]
\centering
\includegraphics [width=12cm,height=7cm]{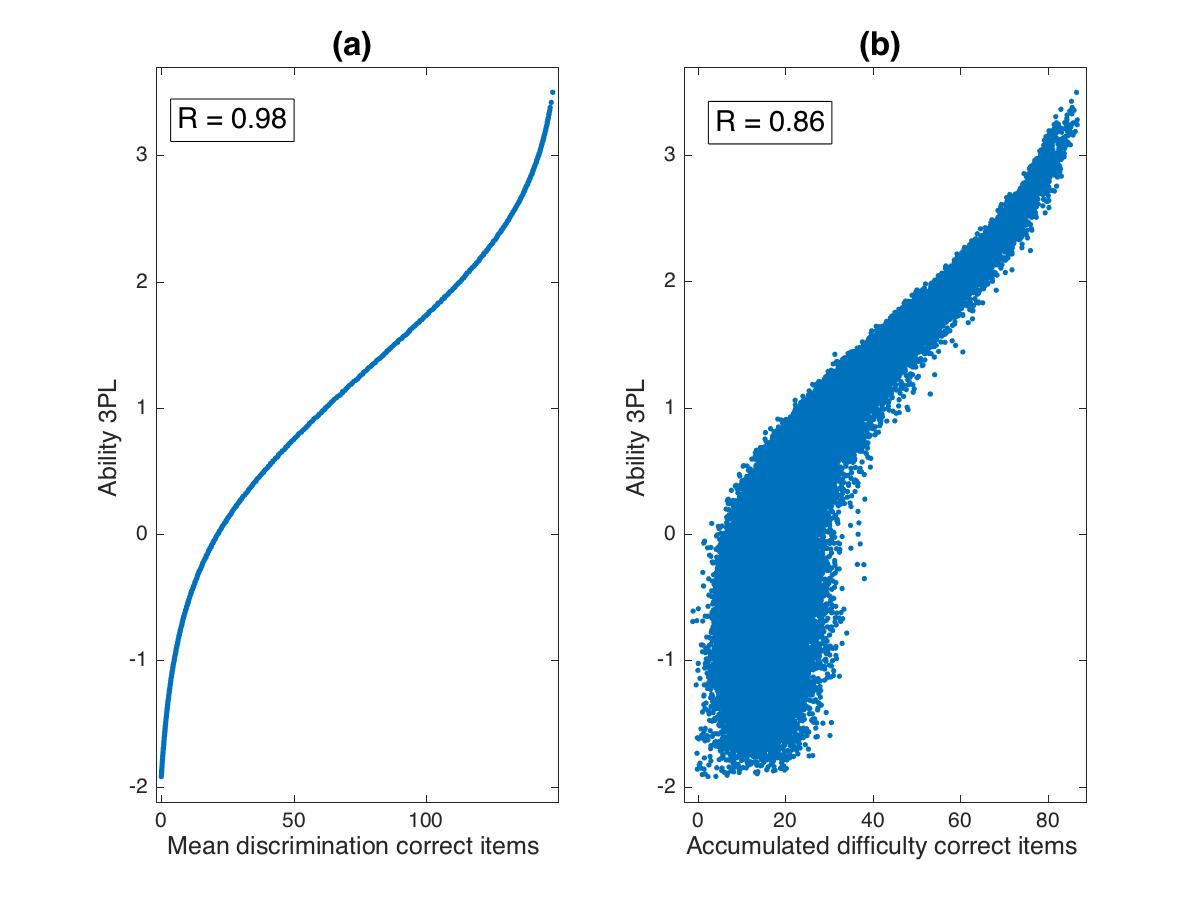}
\vspace{-0.5cm}
\caption{Accumulated discrimination and difficulty v/s ability 3PL}
\label{fig:8}
\end{figure}
\clearpage

\begin{figure}[h]
\centering
\includegraphics [width=12cm,height=8cm]{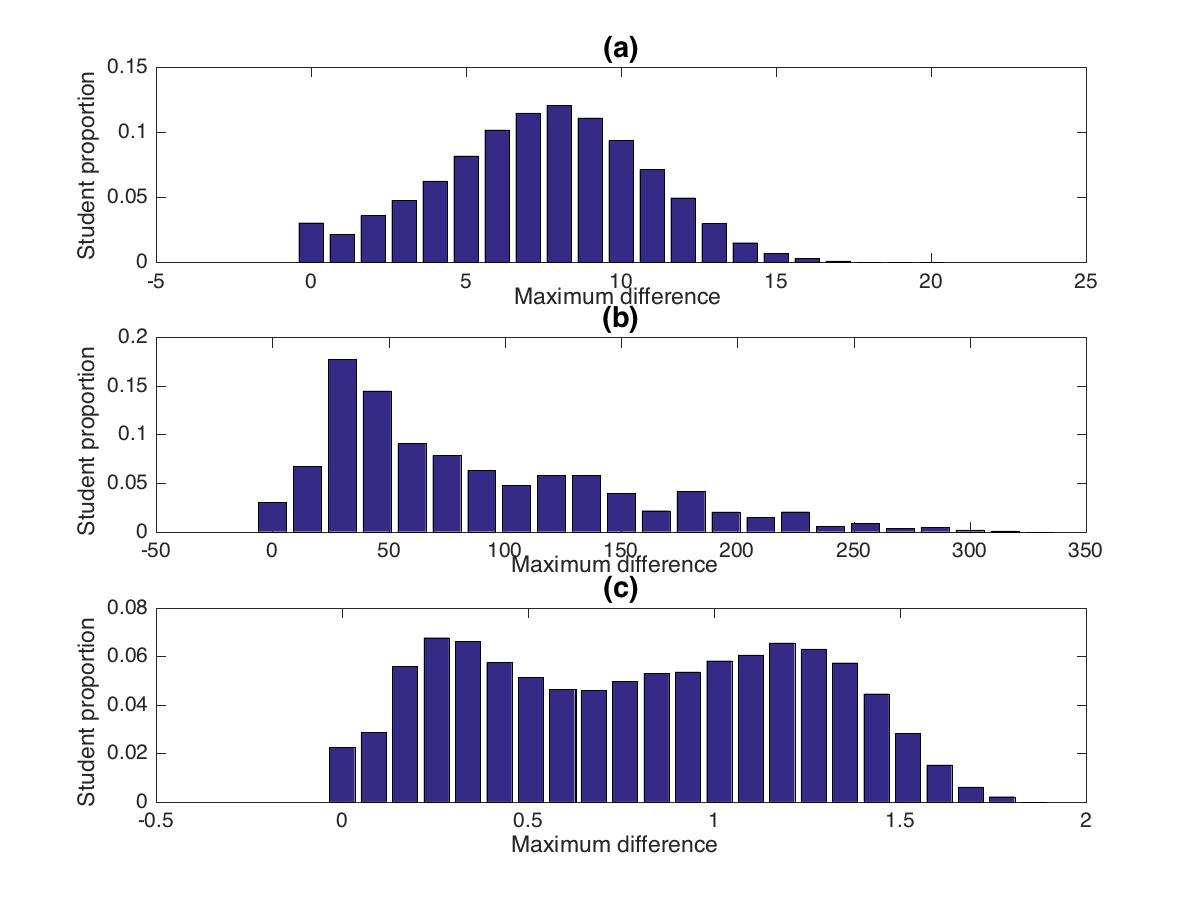}
\vspace{-0.5cm}
\caption{Difference distribution in 3PL}
\label{fig:9}
\end{figure}

\begin{figure}[h]
\includegraphics [width=15.3cm,height=12cm]{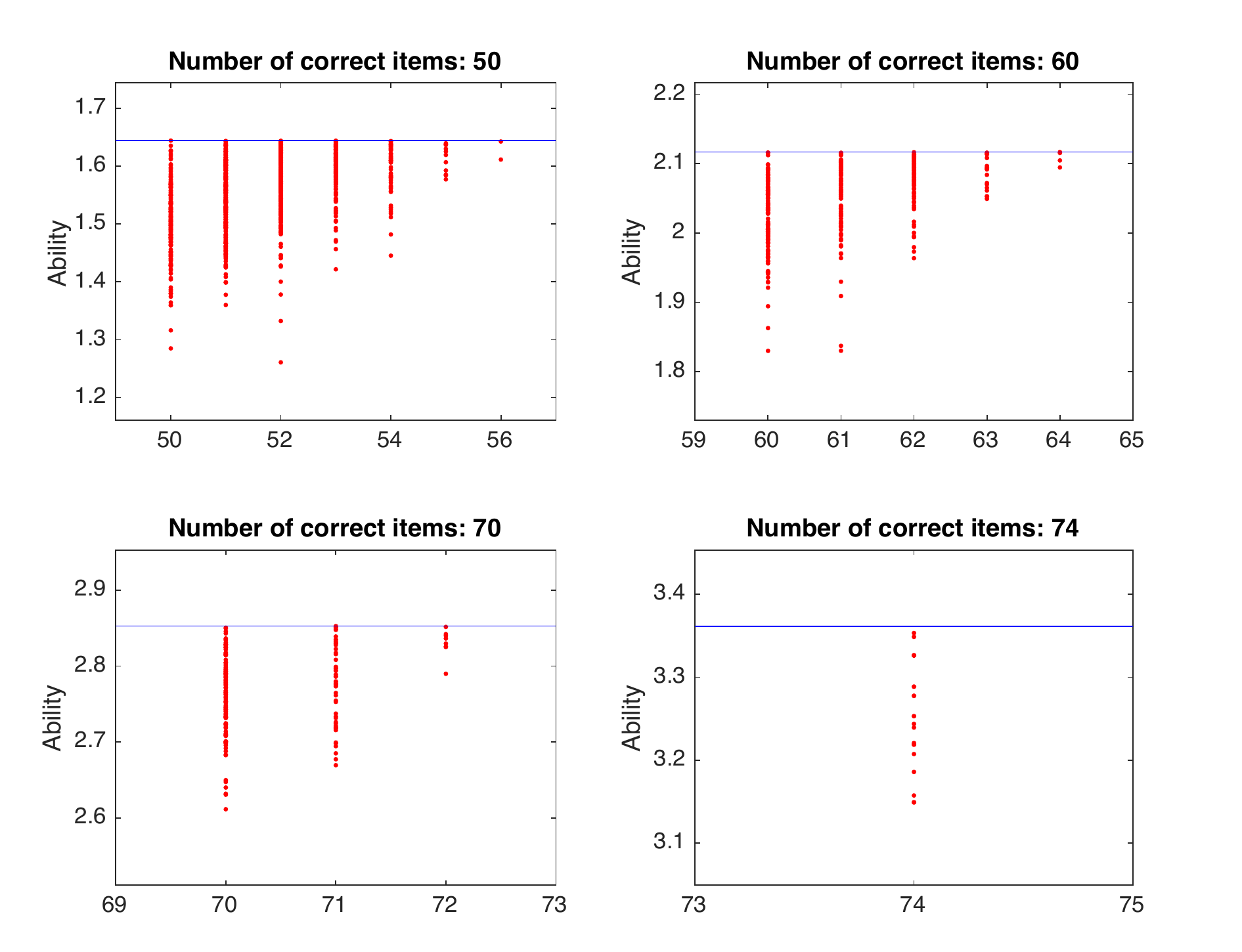}
\vspace{-1cm}
\caption{Cases 50, 60, 70 and 74 correct items, 3PL model}
\label{fig:9.5}
\end{figure}

\end{document}